\documentclass{article}

\usepackage{arxiv}

\usepackage[utf8]{inputenc} 
\usepackage[T1]{fontenc}    
\usepackage{hyperref}       
\usepackage{url}            
\usepackage{booktabs}       
\usepackage{amsfonts}       
\usepackage{nicefrac}       
\usepackage{microtype}      
\usepackage{lipsum}		
\usepackage{graphicx}
\usepackage{doi}
\usepackage[square,numbers]{natbib}
\usepackage{multirow}
\usepackage{array}
\newcolumntype{P}[1]{>{\centering\arraybackslash}p{#1}}

\title{Comparative Study of Machine Learning Models for Stock Price Prediction}

\author{{Oğulcan E. Örsel} \\
	Department of Electrical \& Computer Engineering\\
	University of Illinois Urbana-Champaign\\
	\texttt{oorsel2@illinois.edu} \\
	\And
	{Sasha S. Yamada} \\
	Department of Electrical \& Computer Engineering\\
	University of Illinois Urbana-Champaign\\
	\texttt{sashasy2@illinois.edu} \\
}

\begin{document}
\maketitle

\begin{abstract}
In this work, we apply machine learning techniques to historical stock prices to forecast future prices. To achieve this, we use recursive approaches that are appropriate for handling time series data. In particular, we apply a linear Kalman filter and different varieties of long short-term memory (LSTM) architectures to historical stock prices over a 10-year range (1/1/2011 -- 1/1/2021). We quantify the results of these models by computing the error of the predicted values versus the historical (i.e., ``true'') values of each stock. We find that of the algorithms we investigated, a simple linear Kalman filter can predict the next-day value of stocks with low-volatility (e.g., Microsoft, 'MSFT') surprisingly well. However, in the case of high-volatility stocks (e.g., Tesla, 'TSLA') the more complex LSTM algorithms significantly outperform the Kalman filter. Our results show that we can classify different types of stocks (i.e., clustering) and then train an LSTM for each stock type. This method could be used to automate portfolio generation for a target return rate.

\end{abstract}


\section{Introduction}
There is obvious financial incentive in applying analytical tools to stock market data. Simultaneously, it has long been debated whether an individual (or computer algorithm) can outperform the stock market itself. A famous example of this on-going debate is a million-dollar bet made between Warren Buffet and a hedge fund company \cite{NPR_bet}. Both parties chose different investment strategies over a 10-year period, with Buffet investing in a market index fund and the hedge fund investing in a hand-picked proprietary portfolio of companies. In the end, the index fund chosen by Buffet outperformed the hedge fund portfolio, proving that it is extremely challenging (if not impossible) to beat the stock market. Undeterred by this anecdote, we seek to forecast future stock prices for our CS 545 final project.

\subsection{Stock market intuition}
One of the earliest theories about the behavior of stock prices is that they follow a ``random walk,'' implying that the price of a given stock has equal probability of increasing or decreasing the next day \cite{random_walk}. This theory has remained divisive amongst academics (e.g., economists, statisticians) and financial professionals (e.g., forecasters, financial fund managers) who tend to disagree about whether the stock market can be beat.
More recently, a revised version of this theory known as the Efficient Market Hypothesis (EMH) has been adopted. The EMH reaches a similar conclusion as the random walk hypothesis--- that ultimately future stock prices cannot be meaningfully predicted---but attributes this uncertainty to the complex nature of the stock market and inaccuracies in the valuation of stocks \cite{emh}.

\begin{figure} [!htp]
	\centering
	\includegraphics[width=\textwidth]{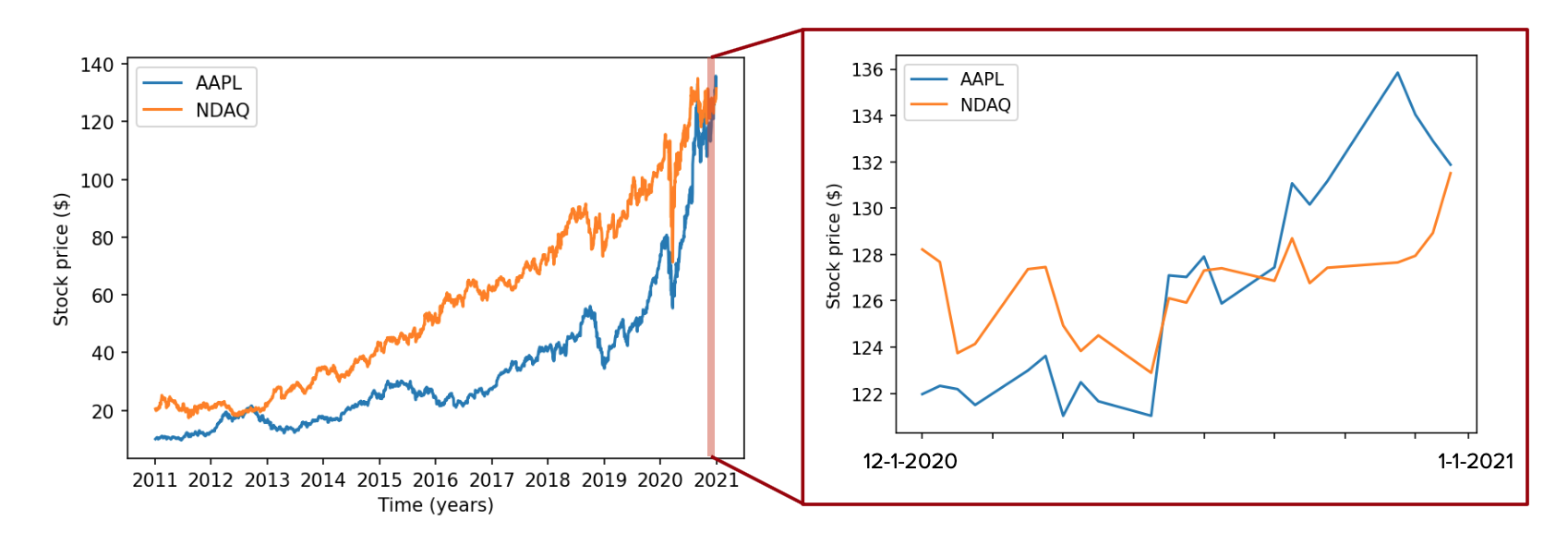}
	\caption{Illustration of long-term and short-term stock market price behavior. In the long-term, stock prices tend upward (but not necessarily in a linear fashion). In the short-term, stock-prices follow a random walk.}
	\label{fig:long_short}
\end{figure}

Instead of choosing a theory to side with, we adopt a more naive view of stock prices. In Fig.~\ref{fig:long_short}, we show the historical stock prices of Apple (AAPL) and a common market index, the NASDAQ (NDAQ), over a long-term and short-term range of time. From observing the trends of these stocks, we use the following intuition to guide our predictions: (1) stock prices tend upward over long periods of time, e.g., years, (2) stock-prices follow a random walk model in the short-term, e.g., days, months. It is important to note that these observations apply to stocks that compose an appreciable share of the overall stock market (i.e., publicly traded companies that influence the S\&P 500 index). We also assume that daily opening/closing stock prices represent the true value of each stock and do not adjust for discrepancies in true valuation. 

\subsection{Objectives}
In this report, we investigate whether machine learning techniques can successfully predict future stock prices. 
We choose to apply algorithms that are appropriate for time-series data, such as Kalman filtering \cite{kalman} and long-short term memory architectures \cite{predict_alg,LSTM_deeplearn}, and compare their performance.
Reliable forecasting results could be extended to potentially improve automatic asset allocation in financial portfolios, or construct a predictive day-trading algorithm.  

\section{Methods}
For this project, we used the library \textit{yfinance} \cite{pypi} to obtain our data set of historical stock prices. We choose to focus on individual stocks from the technology sector (e.g., MSFT, AAPL, TSLA) and well-known market indices (e.g., NASDAQ, S$\&$P 500). We use TSLA as an example of a volatile stock, and MSFT as an example of a non-volatile stock. For Kalman filter, we did not applied any scaling to our data, however for our LSTM models we first scaled our data between 1 and 0. Due to our memory structure, we also generated 3D time series before training the model. After our prediction, we rescaled our output to compare with the stock data.

\subsection{Kalman filter}
Kalman filtering is a recursive algorithm that is used in many tracking/forecasting applications (e.g., GPS, robotics). In its simplest form, a Kalman filter balances measurement uncertainty and prediction uncertainty to provide a more accurate estimate of the current state than either metric on its own. Since we treat the short-term behavior of stock prices as a random walk, we implemented a linear Kalman filter with the following assumptions: (1) the variance of the previously measured state, i.e., the previous day's stock price, is very small, and (2) the variance of the predicted state will be proportional to the local historical variance of the current state. For the model that we implemented, we found using the local variance of the previous three days worked well. 

\subsection{Long short-term memory (LSTM)}

\begin{figure}[hbtp]

    \centering
    \includegraphics[width=0.85\textwidth]{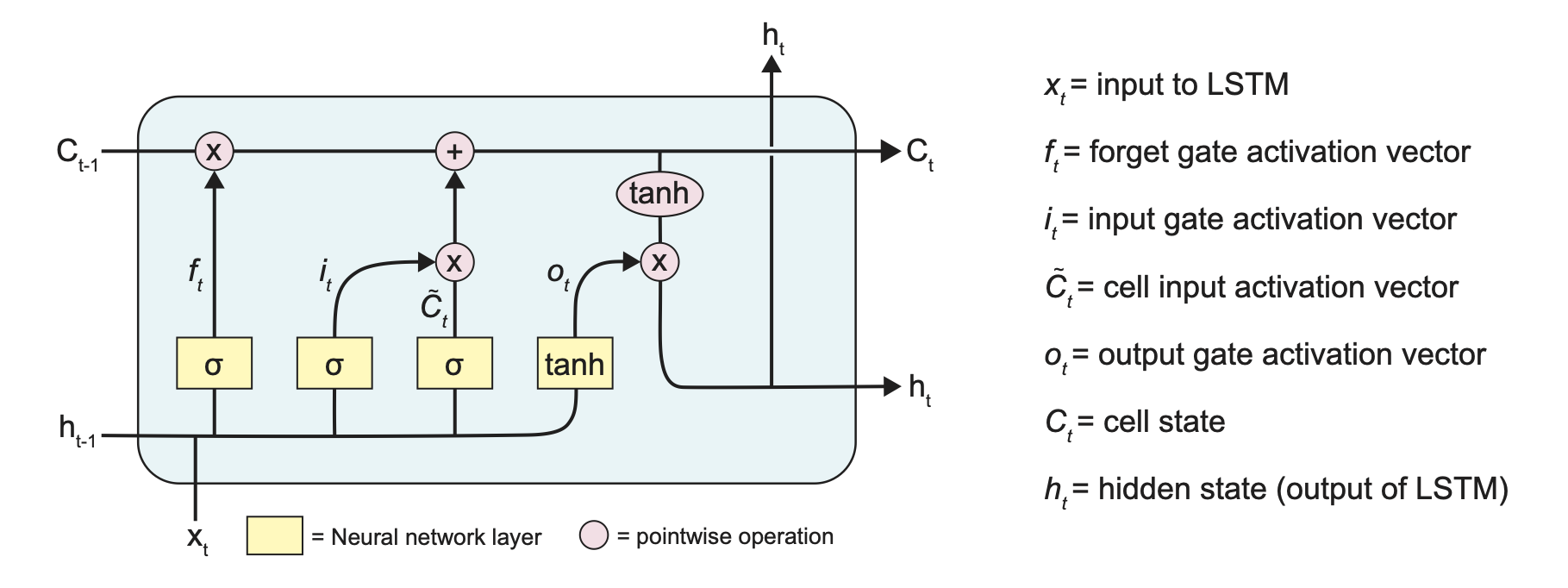}
    \caption{LSTM structure, inspired by \cite{LSTM_intro}.}
    \label{fig:LSTM}

\end{figure}

Long short-term memory networks (LSTMs) are unique recurrent neural networks (RNNs) that can learn long term dependencies. They were introduced by Hochreiter $\&$ Schmidhuber (1997) \cite{LSTM_intro}, and improved by many people. The main advantage of the LSTM architecture is that its memory structure can choose the duration of the learning procedure in a way that eliminates long-term biases. The main structure of a generic LSTM network is given in Fig.~\ref{fig:LSTM}.

Referring to Fig.~\ref{fig:LSTM}, we observe that there are three main variables: C\textsubscript{t}, h\textsubscript{t} and x\textsubscript{t}. These variables represent the memory (C\textsubscript{t}), the output (h\textsubscript{t}), and the input (x\textsubscript{t}) of the LSTM network. There are also three primary processes performed within the structure. The first process determines the structure of the memory unit (f\textsubscript{t}) for the following step via element-wise multiplication. The previous output and the current input determine this structure by using a neural network layer (usually called the forget gate). The second process is to evaluate the cell state, which is affected by the input gate, current input and previous output, and is determined using a neural network. This unit mixes the current input with the previous memory and prepares the cell state to enter the next gate. Finally, we have the last unit which determines the current output with the previous output, current input and the cell state with a neural network.

There are various LSTM architectures which applications \cite{types_LSTM}. In our work, we implemented a single layer LSTM, a stacked LSTM, a bidirectional LSTM and a convolutional neural network (CNN)-LSTM. We chose to compare their performance in terms of well known metrics (root-mean square error (RMSE), mean absolute error (MAE), and R\textsuperscript{2} value). While all models use the LSTM framework, there are slight differences between them. For example, the stacked LSTM uses more than one layer, the bidirectional LSTM learns the time series in both directions, and the CNN-LSTM uses convolution neural networks to enhance the feature extraction process.

\section{Experiments}
In this section we discuss the forecasting experiments that we performed. We first compared the results of various models by computing the root-mean square error (RMSE), mean absolute error (MAE), and R\textsuperscript{2} value for each model. The smaller the error metrics (RMSE and MAE), the better the performance. Since the R\textsuperscript{2} value assesses the fit of a data set to a given regression, numbers closer to one are better. We next selected the model with the best performance and tested whether it would still yield reasonable results if applied to a market index with similar behavior as the training stocks.

For our LSTM models, we optimized them so that we got quality performance while keeping the model simple. For instance, our single layer and bidirectional LSTMs consist of 64 nodes and with a memory size of three. Moreover, our double LSTM structure consists of two consecutive 64 nodes with a memory size of 3. Finally, our CNN-LSTM performs a 1D convolution with a size of 64 steps and also uses an LSTM network with 64 nodes. After each LSTM was individually optimized, we learned two different types of stocks (volatile $\&$ non-volatile) and compared the performance of the all models.

\subsection{Comparison of models}

We calculated the RMSE, MAE and R\textsuperscript{2} values for TSLA stock under different models and the results are given in Table 1. We found that the CNN-LSTM performed very well at predicting future stock values. We note that the performance of the other LSTM models was similarly impressive. We believe that this is a result of the effectiveness of the CNN-LSTM's feature extraction ability compared with other algorithms. The worst performing LSTM model was the dual layer LSTM. We attribute this to increased lag caused by the double layer structure. 

\begin{figure}[htp]
    \centering
    \includegraphics[width=4in]{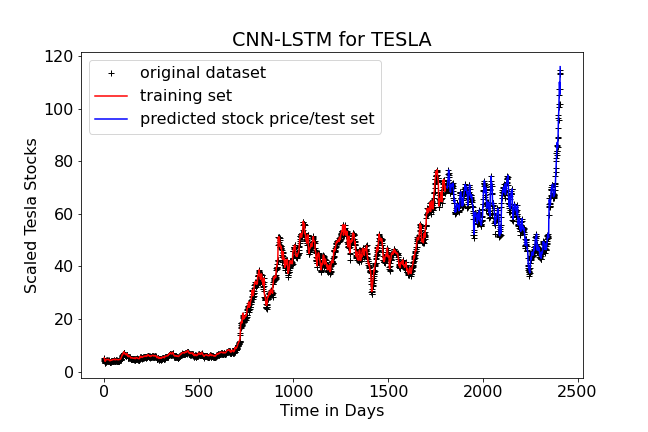}
    \caption{CNN-LSTM fit for TSLA stock. The training data set consists of 7.5 years of historical prices, and the testing set used to predict next-day prices consists of 2.5 years of prices.}
    \label{fig:TSLA}
\end{figure}

One important takeaway is that the performance of the Kalman filter is greatly improved with the inclusion of a more complex LSTM model. This validated our earlier reasoning that including a more complex algorithm would improve accuracy of our forecasting. From these tests, we can conclude that it is better to use machine learning algorithms that are effective at extracting features (such as bidirectional-LSTM and CNN-LSTM) for stocks with volatile prices. Finally, in Fig.~\ref{fig:TSLA} we plot the results of applying our CNN-LSTM model, and highlight the test data, training data, and the predicted prices.

\begin{centering}
\begin{table}[htp]
\centering
\caption{\label{tab:table-name} Performance of the algorithms for Tesla stock.}
\begin{tabular}{ |P{3cm}||P{3.5cm}|P{3.5cm}|P{3.5cm}|  }

 \hline
 Algorithms & Root Mean Square Error & Mean Absolute Error & R Value\\
 \hline
 Kalman Filter & 6.55 & 55.98 & 0.76\\
 Single Layer LSTM & 2.44 & 1.74 &  0.95\\
 Dual Layer LSTM & 2.66 & 1.84 &  0.94\\
 Bidirectional LSTM  & 2.48 & 1.79 &  0.95\\
 CNN-LSTM & 2.20 & 1.54 & 0.96\\
 \hline

\end{tabular}
\end{table}
\end{centering}

\begin{figure}[h]
    \centering
    \includegraphics[width=4in]{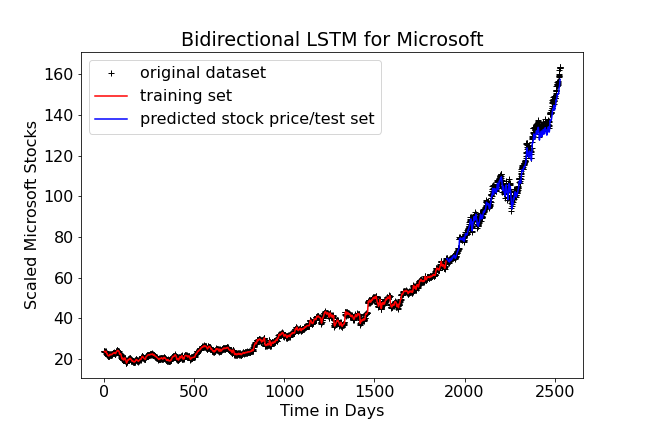}
    \caption{Bidirectional LSTM fit for MSFT stock. The training data set consists of 7.5 years of historical prices, and the testing set used to predict next-day prices consists of 2.5 years of prices.}
    \label{fig:MSFT}
\end{figure}

Similarly, we calculated our metrics RMSE, MAE and R\textsuperscript{2} for the MSFT stock as well as given in Table 2. We found that the bidirectional LSTM and Kalman filter performed very well. For the LSTM, we believe that this is a result of the effectiveness of the bidirectional-LSTM's feature extraction ability compared to other algorithms. Similar to the TSLA case, the worst performing algorithm was the dual layer LSTM---it was even worse than the Kalman filter. That lead us to conclude that the dual layer LSTM is not the best option for stock price prediction, as single-layer structures consistently produced better results. We also note that for non-volatile stocks such as MSFT, the Kalman filter performs really well and complex LSTM algorithms are not necessary to produce high accuracy predictions. Finally, in Fig.~\ref{fig:MSFT} we plot the results of our bidirectional-LSTM model, and highlight the test data, training data and the predicted prices.

\begin{centering}
\begin{table}[h]
\centering
\caption{\label{tab:table-name} Performance of the algorithms for MSFT stock.}
\begin{tabular}{ |P{3cm}||P{3.5cm}|P{3.5cm}|P{3.5cm}|  }

 \hline
 Algorithms & Root Mean Square Error & Mean Absolute Error & R Value\\
 \hline
 Kalman Filter   & 4.78    & 1.19 &  0.99\\
 Single Layer LSTM&   3.51  & 2.64   &0.98\\
 Dual Layer LSTM& 24.66 & 19.97 &  0.63\\
 Bidirectional LSTM   &2.81 & 2.14&  0.99\\
 CNN-LSTM &  4.32  & 3.49 & 0.97\\
 \hline

\end{tabular}

\end{table}
\end{centering}
\subsection{Applying trained LSTM to different stocks}
\begin{figure}[h]
    \centering
    \includegraphics[width=\textwidth]{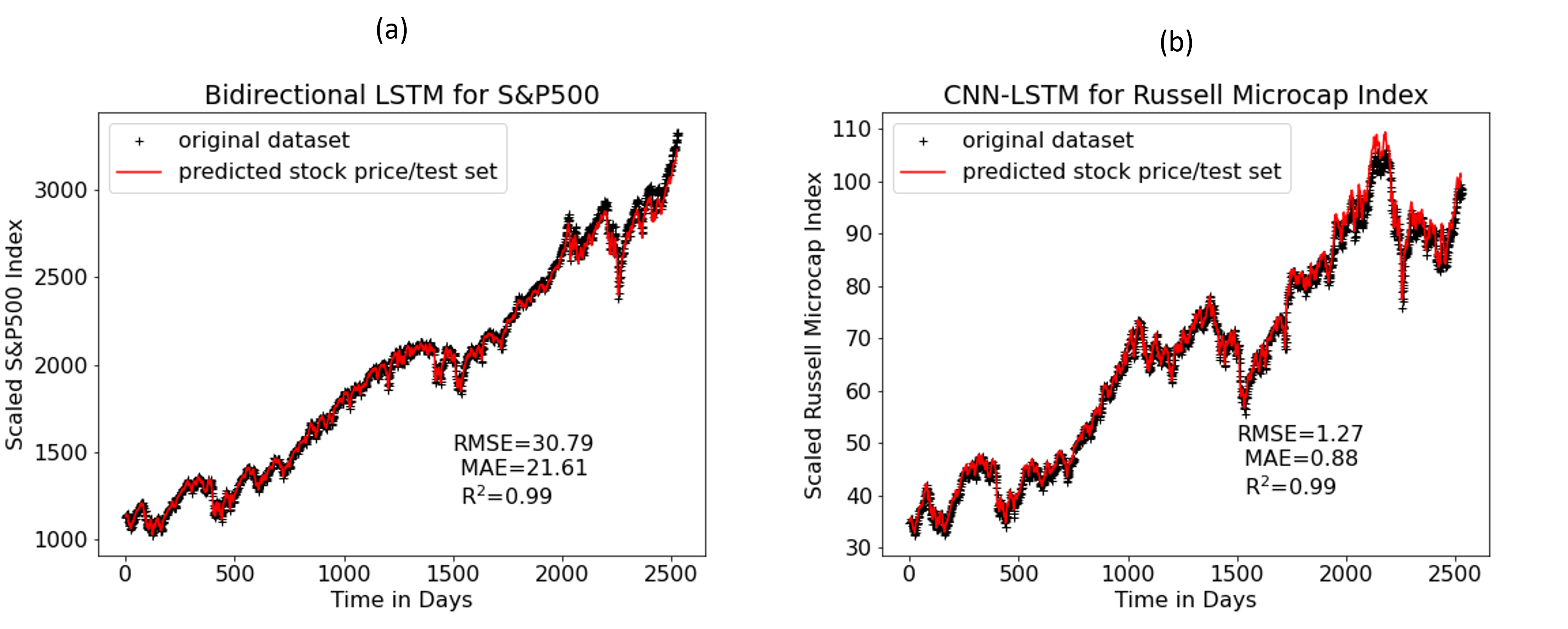}
    \caption{Extension of our best-performing LSTM models to two different market indexes.}
    \label{fig:extension}
\end{figure}

As a final test, we applied our champion models (bidirectional and CNN LSTM) to market index data. As a non-volatile model, we tested the S$\&$P 500, and as a volatile model we tested Russel microcap index. These market indexes were selected to reflect our previous candidate stocks (TSLA and MSFT). The prediction results are given in Fig.\ref{fig:extension} with their performance metrics.

Referring to Fig.~\ref{fig:extension}, we can clearly see that both models fit the data very well with an R$^{2}$ values reaching 0.99. This indicates that our models can be used to predict volatile or non-volatile types of stocks. Furthermore, it also shows that it is not necessary to re-train an LSTM for each individual stock, as stocks/market indexes with similar levels of volatility can be accurately forecasted with previously trained LSTMs. That provides us with a path to simplify the stock market problem. An extension of this study could be made to classify different types of stocks (i.e., clustering) and then train an LSTM for each stock type. This method could be used to automate portfolio generation for a target return rate.

\section{Conclusion}
The goal of this project was to apply machine learning techniques to stock price forecasting. We were able to successfully implement a linear Kalman filter and several more involved LSTM models for this purpose. We found that the accuracy of each model is substantially affected by the volatility of the stock being forecasted. For low-volatility stocks, a linear Kalman filter can predict next-day prices with very reasonable accuracy. However, this error increases significantly for more volatile stocks, making LSTM architectures a much more suitable choice. We also tested our best performing LSTM models (a bidirectional LSTM for low-volatility stocks, and a CNN-LSTM for high-volatility stocks) on well-known market indexes. Without retraining the LSTM models, we find that they can predict next-day stock prices with reasonable accuracy. We believe that the results of our study could be extended to construct an automated stock portfolio generator, or alternatively a day-trading algorithm that suggests mechanical trading operations. Overall, we have found these results promising and intend to submit an abstract to the 17th Coordinated Science Laboratory Student Conference that will take place in February 2022. 

\bibliographystyle{unsrtnat}
\setcitestyle{square,numbers,comma}
\bibliography{references}  

\end{document}